# Micro-Synchrophasors for Power Distribution Monitoring, a Technology Review

Reza Arghandeh

Dept. of Electrical and Computer Engineering, Center for Advanced Power Systems, Florida State University, Tallahassee, USA,

arghandehr@gmail.com

*Abstract:*

The smart grid revolution is creating a paradigm shift in distribution networks that is marked by new, significant intermittencies and uncertainties in power supply and demand. These developments include the dramatic increase in the adoption of distributed energy resources (DER), electric vehicles, energy storage, and controllable loads. This transformation imposes new challenges on existing distribution infrastructure and system operations for stockholders, engineers, operators and customers. Unfortunately, distribution networks historically lag behind transmission networks in terms of observability, measurement accuracy, and data granularity. The changes in the operation of the electric grid dramatically increase the need for tools to monitor and manage distribution networks in a fast, reliable and accurate fashion. This paper describes the development process of a network of high-precision micro phasor measurement units or µPMUs, beginning with an overview of the µPMU technology that provides synchronous measurements of voltage phase angles, or synchrophasors. Next, the µPMU network and communications infrastructure are discussed, followed by an analysis of potential diagnostic and control applications of µPMU data in the electric grid at the distribution level.

## I. INTRODUCTION

Historically, with mostly radial power distribution and one-way power flow, it was only necessary to evaluate the envelope of design conditions, e.g. peak loads or fault currents, rather than continually observe the operating state. The lack of visibility on distribution systems follows from simple economics: there has never been a pressing need to justify extensive investment in sensing equipment and communications. Even with the growing need for monitoring capabilities, the costs must be far lower to make a business case for measurement devices on a distribution circuit compared with the transmission setting [1]. But the growth of distributed energy resources (DERs) introduces variability, uncertainty, and opportunities to recruit diverse resources for grid services. The presence of multiple generation resources on each feeder has complex impacts on the circuit behavior that can be observed in the variation of voltage and current phase angles [2]. DERs also represent a new menu of options for grid support functions, such as volt-VAR optimization, adaptive protection, efficiency enhancement, ancillary services or even intentional islanding. This suggests a need for more refined measurement, given both the challenge of managing increased variability and uncertainty and the opportunity to recruit diverse resources for services in a more flexible grid. This paper addresses how the direct measurement of voltage and current phasor values (magnitude and phase angle) might enable new strategies for managing distribution networks with diverse, active components. Specifically, it discusses high-precision micro phasor measurement units (µPMUs), that are tailored to the particular requirements of power distribution and can support a range of diagnostic and control applications, from solving known problems to opening as-yet-unexplored possibilities.

## II. MICRO-SYNCHROPHASOR MEASUREMENTS

A µPMU is a high-precision power disturbance recorder adapted for making voltage phase angle or synchrophasor measurements, capable of storing, analyzing and communicating data live. Hardware and software for the µPMU was initially developed by Power Sensors Limited (PSL) with the University of California and Lawrence Berkeley National

Lab (LBNL), in a collaborative project funded by the U.S. Department of Energy ARPA-E beginning in 2013, to address the need for tools to better observe, understand and manage the grid at the distribution scale [3, 4]. Author was with UC Berkeley and contributed in this project during 2013-2015. A key technological innovation is the precise time stamping of measurements via GPS to allow the comparison of voltage phase angle δ (i.e., the precise timing of the voltage waveform) at different locations. On the software side, UC Berkeley researchers developed a new computational framework termed the Berkeley Tree Database (BTrDB) to manage large, high-density data streams with nanosecond time stamping and online capabilities that obviate the need for phasor data concentrators [Andersen DISTIL paper]. The project team (including the authors) collaborated with several electric utilities to install a network of µPMUs on various distribution circuits, study the early data and begin to develop practical applications for this new type of information. Several other research projects have since begun to employ µPMU measurements in the distribution system context with a range of application foci, including the integration of intermittent renewable generation, distributed control, and cyber-security of the electric grid.

An overarching research question is how the direct measurement of voltage and current phasor values can address both known and poorly understood problems, such as dynamic behaviors on the distribution grid, that are becoming increasingly relevant in the context of smart grid evolution. The underlying physics provides that measurement of voltage phase angles can serve as a proxy for power flow between two points on an a.c. network, which varies mainly with the voltage angle difference δ between those two points. When the line impedance is mainly inductive, as it is in the transmission context, power flow is approximated by:

$$P_{12} \approx \frac{V_1 V_2}{X} \sin \delta \qquad (1)$$

where $X$ is the inductive reactance of the line, and $V_1$ and $V_2$ are the voltage magnitudes. Because real and reactive power flow throughout an a.c. network is uniquely determined by voltage phasors (i.e., magnitudes and phase angles) at each node, the voltage phasors are considered *state variables*.

Unlike transmission conductors, distribution lines are typically characterized by a significant electrical resistance R along with the reactance X (which is dominated by series inductance but may also include a non-negligible shunt capacitance, especially for underground cables). Therefore, the relationship between voltage magnitudes, angles and real and reactive power flows is much less amenable to convenient approximations like Equation (1). In fact, distribution system modeling often requires explicitly considering the imbalanced interaction among the three phases, yielding a much more awkward set of complex equations to describe power flow. Nevertheless, the voltage phasor is the state variable that in theory captures all the necessary information about the behavior of the a.c. system, at and above the time scale of one cycle. The question, then, is whether a sufficiently accurate phasor measurement can be made in practice, and appropriate analytics be developed, to support a given application in a particular context, and whether such a µPMU-based approach to grid diagnostics or control is efficient and economical compared to the alternatives.

Through empirical measurements in conjunction with modeling and analysis of distribution circuits, our ARPA-E project has found early evidence for the usefulness of phase angle as a measured quantity, while also identifying challenges in working with µPMU data. It has validated the performance of µPMU hardware and produced open-source software that could enable a new management approach for distribution systems, particularly in the presence of significant renewable penetration.

### III. Synchrophasor Technology Overview

PMUs were first introduced by Virginia Tech University researchers in the early 80s [5]. They are a mature technology in transmission networks, and are used in a variety of applications[6]. The major blackout of 2003 in the Northeastern United States revealed the emerging need for time-synchronized measurement in cascading failure diagnostics and prognostics. Since then, PMUs have been a cornerstone of wide-area measurement systems (WAMS). However, synchrophasors are not used widely to observe power distribution networks. Although "distribution" PMUs may be deployed at distribution substations, their voltage angle measurements are usually referenced against angles elsewhere on the transmission grid, not the distribution feeder. In contrast, the purpose of micro-synchrophasors is to compare voltage angles at different points on distribution circuits, all behind the substation. These "true" distribution applications are much more challenging, in several respects:

1) Because power flows are small, voltage angle differences on a distribution circuit may be two orders of magnitude smaller than those on the transmission network – i.e., tenths of a degree, not tens of degrees. Variations from the steady state must therefore be measured on an even finer scale, down to tens of millidegrees.

2) Distribution systems measurements are much noisier than transmission systems, where variability tends to be better smoothed by statistical aggregation (e.g. of time-varying loads). Since high-resolution measurements at the medium-voltage distribution level are still rare, there is not yet a solid body of empirical knowledge about extracting the relevant signal from the noise at different time scales in distribution system measurements.

3) Due to the short distance between network components and the higher density of power electronics and controlled devices in distribution networks, the high-resolution measurement noise includes harmonics and small transients associated with nearby devices or switching operations on a circuit. For this reason, we expect that it will prove useful to combine µPMU data (which by definition applies to a full a.c. cycle) with power quality measurements that reveal harmonics on the sub-cycle scale.

3) The costs must be far lower to make a business case for the installation of multiple µPMUs on a distribution circuit, as compared to the clear economic rationale for PMU installation in a transmission setting.

4) The ratio of available empirical data points to the number of network nodes is much smaller in distribution than in transmission, given that each service transformer is effectively a circuit node and that smart meter data are not generally communicated and available in real time. Consequently, it is much more difficult to perform rigorous state estimation for the entire network.

Fig. 1 illustrates µPMU capabilities and measurements on a logarithmic time scale. Some of these placements are approximate and are being continuously refined as we better understand the rates at which it is practical and useful to report certain measurements.

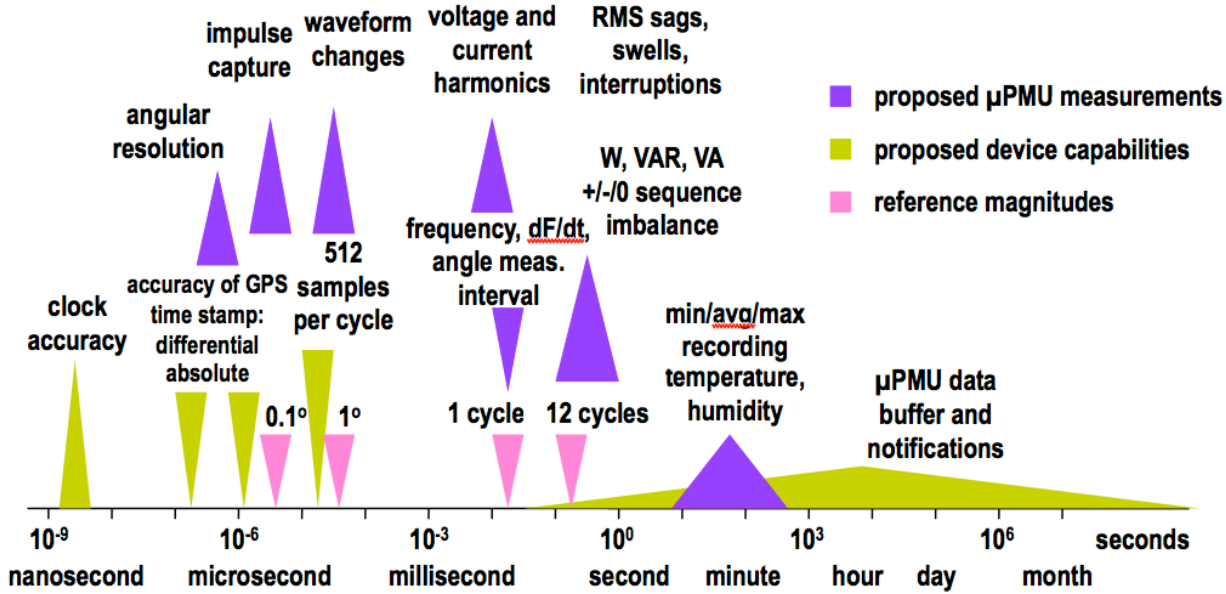

Figure 1. Time Scale for µPMU Performance

The µPMU technology can discern angle separations as small as 0.01°, and it can be toggled to operate contextualizes these data within high-resolution power quality recordings [4]. Its components include a power quality measurement (PQube) instrument that contains the measurement, recording, and communication functionality, a remotely-mounted micro GPS receiver, and a power supply with battery backup (all easily fit inside a small breadbox). µPMUs continuously sample a.c. voltage and current waveforms at 120 samples per cycle. They record a wide range of detailed power quality measurements and environmental conditions, either continuously or as triggered by disturbances. All µPMU data are stored in files on a Secure Digital (SD) card which allows for months of data storage and assures data recovery if communications are lost during power system events. Internal Ethernet support includes a PQube web server File Transfer Protocol (FTP), and a universal email client.

There are several common challenges faced in installing and operating µPMUs, one of which is the need to electrically isolate the GPS receiver to resist lightning strikes, while accounting for signal latencies. Another challenge is the computation of phase angle at very high resolution, which can be complicated by the presence of harmonics and by the competing demands of phase-locked versus time-based sampling.

The µPMU device can be connected to single- or three-phase secondary distribution circuits up to 690V (line-to-line) or 400V (line-to-neutral). Measurements can be taken either from standard outlets or from overhead lines through potential transformers (PTs). These PTs are found at distribution substations and can also be added on primary distribution circuits. In any installation, it is necessary to account for the effects of transformers on voltage angle through appropriate analytics. Figure 2 shows µPMU rack mounted and pole mounted installations.

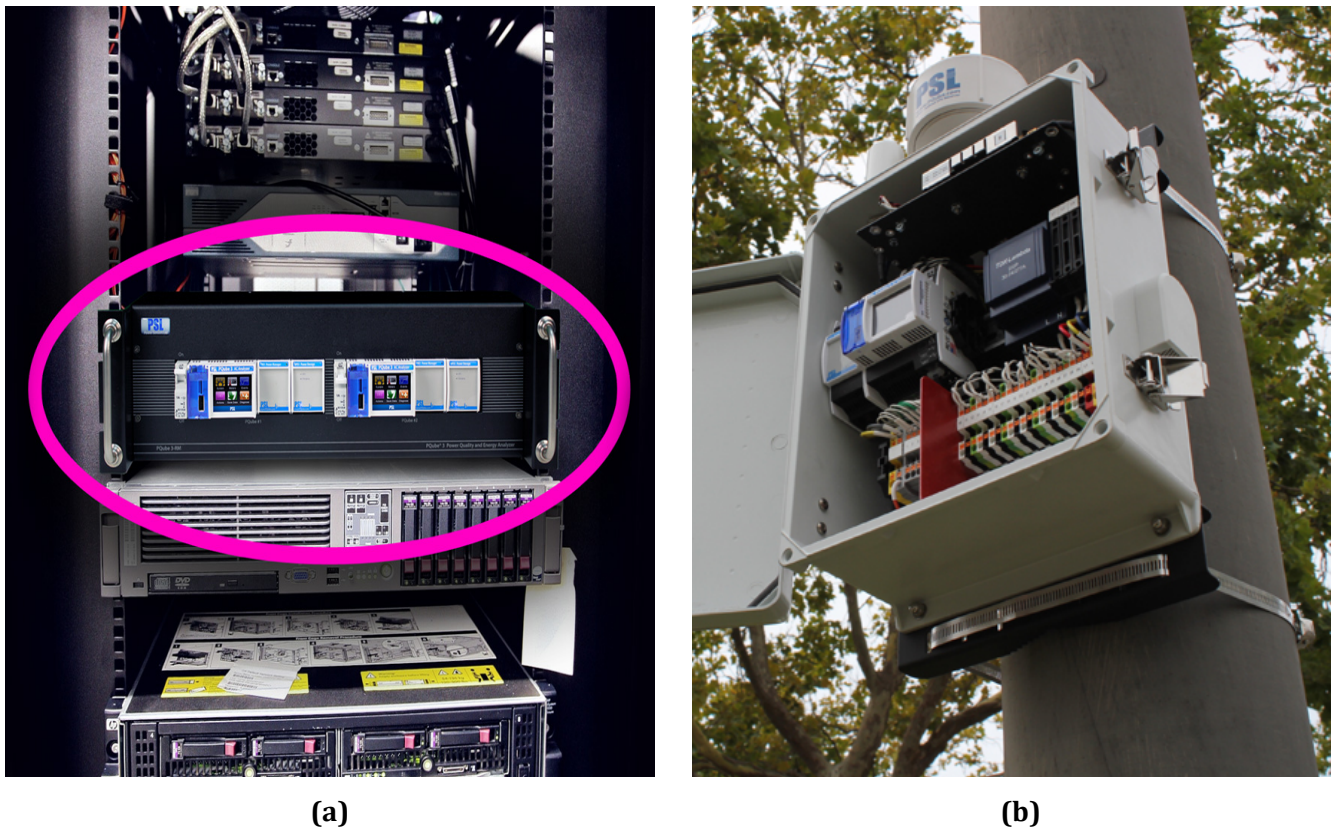

            **(a)**                    **(b)**

Figure 2. PQube3 or µPMU installation types (pictures source: PSL website [4, 7])
 (a) Rack-mount installation
 (b) Pole-mount installation

IV. **A NETWORK OF µPMUS**

The true potential for the use of phase angle data in real-time applications lies in effective networking and data management. Fig. 3 illustrates the deployment concept for the µPMU network [3]. With µPMUs installed at multiple locations throughout a distribution feeder (e.g. the substation, end of feeder, and any key distributed generation facilities), the monitoring networks is intended to support the analysis and operation of an individual feeder, multiple feeders from the same substation, or even contribute to the observation of transmission-level phenomena. A key challenge is accounting for the effects of distribution transformers when measurements are taken on the secondary side, but inferences are to be made about the state of the primary circuit.

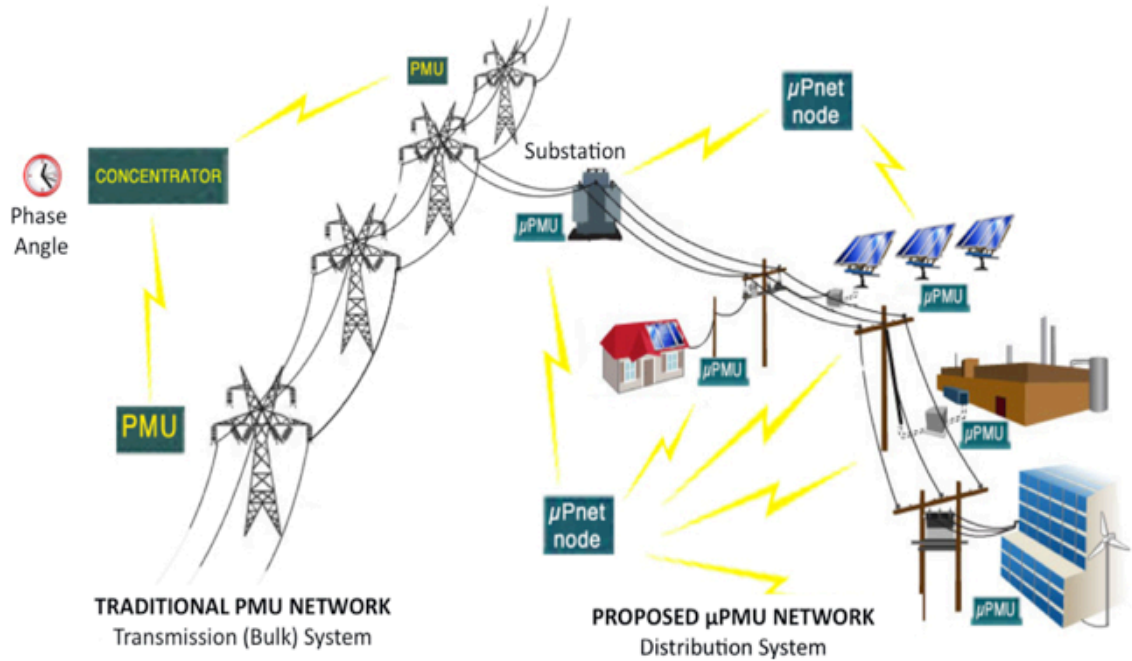

Figure 3. µPMU network concept [3]

The communication interval between a µPMU and a network node may vary as appropriate for the application, e.g., once per cycle, every few seconds, or through reports triggered by anomalous measurements. Depending on the application for which µPMU data are to be used, analyzed data may be displayed in visual format, or forwarded in distilled form to other users. For example, a digest can be sent to the distribution system operator, or a control signal could be sent to selected devices.

However, the main practical barrier for high-resolution monitoring systems is handling the real-time data streams [8]. Data produced by multiple PMUs are typically aggregated in a phasor data concentrator (PDC) which buffers data and provides basic integrity checking, data format checking, and time tag correlations based on few available standards such as the Institute of Electrical and Electronics Engineers (IEEE) C37.118 [9]. The PDC delivers the data over higher bandwidth links to a higher layer data concentrators or back-end data processing resources in control centers. For real-time or quasi-real-time applications, a µPMU uploads its precisely time-stamped measurements through a suitable physical communication layer to a µPnet node, where it is compared against measurements from other µPMUs. The network of µPMUs called the "µPnet". The µPnet is agnostic to the physical communication layer used, although the target speed and bandwidth constrains the selection of the most economical medium, which is the 4G technology. Various data historians are used to warehouse the measurement data streams and various relational and non-relational databases are used to hold small subsets of data and results from the data processing, typically as a part of a vendor patented solution. Very few solutions are available to archive the PMU data sets, especially on the scale of distribution network monitoring (compared to transmission networks, which require fewer PMUs). As the grid system continues to scale up and number of µPMUs increase, the aggregated PMU data will be a high volume, high velocity and high variability data set. It will bring the Big Data era to the distribution networks.

An agile, scalable stream processing infrastructure tailored for µPMU timeseries in developed in the University of California Berkeley, called the Berkeley Tree Database (BTrDB). The µPMU timeseries are delivered into the multi-resolution, versioned time-series data store. The data analysis on streams are described by distillers that perform computation as chunks of data change on input streams and push analysis results onto output data streams. The analysis performed includes basic cleaners on raw data and feed through dataflow operators that may combine time-correlated streams within a µPMU or across µPMUs. A kernel that directly reflects the related mathematics describes the algorithm for each distiller. There is version tracking for the data, the distillers, and the intermediate streams. The tool developed is called the Design and Implementation of a Scalable Synchrophasor Data Processing System (DISTIL) [10] , see figure 4.

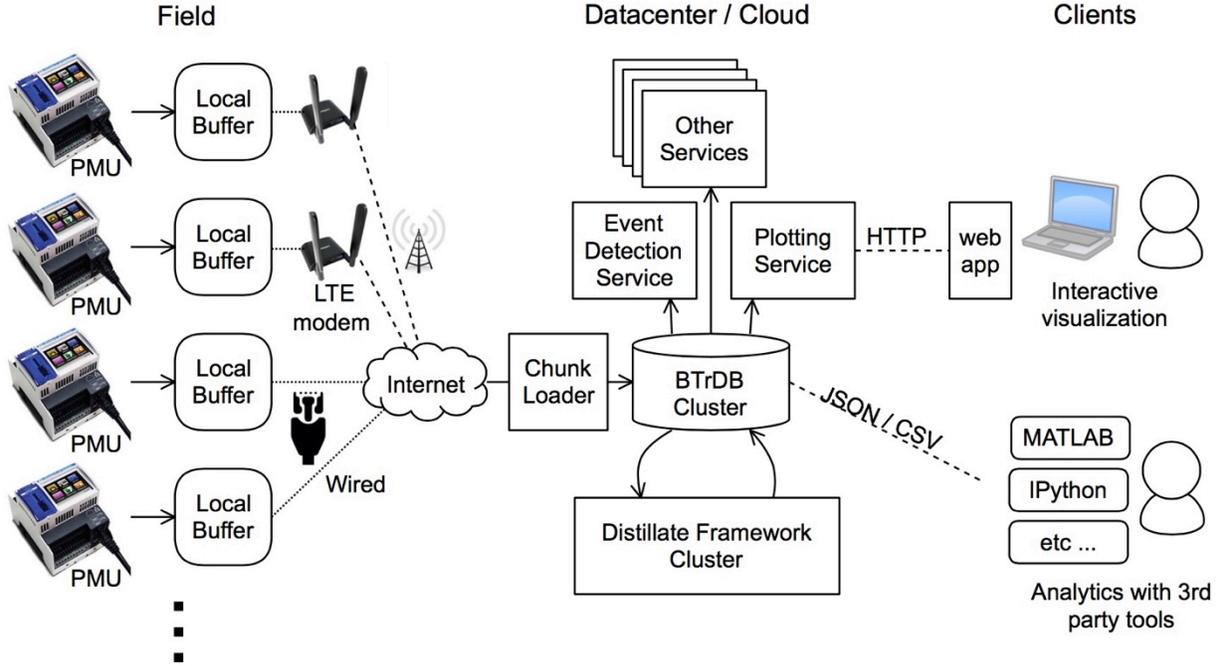

Figure 4. the μPMU data historian, analysis and visualization (DISTIL) architecture [10]

## V. μPMU Data Visualization for Situational Awareness

In complex systems such as power distribution networks containing thousands of components, an agile visualization tool is crucial for operators to observe system states in a quick and intuitive fashion [11]. A capable visualization tool will be even more informative when presenting time-synchronized data streams. The DISTIL shows the system's physical behavior at any given moment in time, as well as its evolution over a range of time scales from milliseconds to months.

An emerging challenge as distribution systems increasingly host diverse and dynamic components such as distributed energy resources and controllable loads is to combine steady-state, transient and dynamic analysis within the same tools to support the network operation [12]. Rather than abstract, unrealistic and suboptimal assumptions for applications development and validation, we need data-driven approaches to learn from our new observations. For example, diagnosing the origin of a voltage variation event as a transmission-level or locally caused event can be corroborated by the magnitude of corresponding voltage and current magnitudes at different locations on the grid, using the DISTIL tool. The primary purpose of DISTIL is to serve as an analytic tool to enable operators to visualize never-before-observable quantities, and develop applications to capture and characterize those events autonomously. In figure 5, variation of voltage and current phasors in different locations at the distribution network level following a transmission level event can be characterized in detail because data and distillates are available simultaneously at sub-cycle temporal resolution and over the course of minutes and hours around the event. In Figure 5, voltage and current phasor data from two different locations (the PSL building in Alameda, CA and a the Soda Hall on University of California Berkeley campus which are 13 miles away) represent spatial-temporal chatterers of the same event. The data and the plot are available for public on Mr. Plotter that developed by UC Berkeley and PSL [13]. The blue lines show voltage magnitudes and red lines show current magnitudes.

It is worth pointing out that measurements, such as those from the μPMU network, represent a vast expansion of visibility into power distribution systems, which are highly diverse, dynamic and thus data-rich.

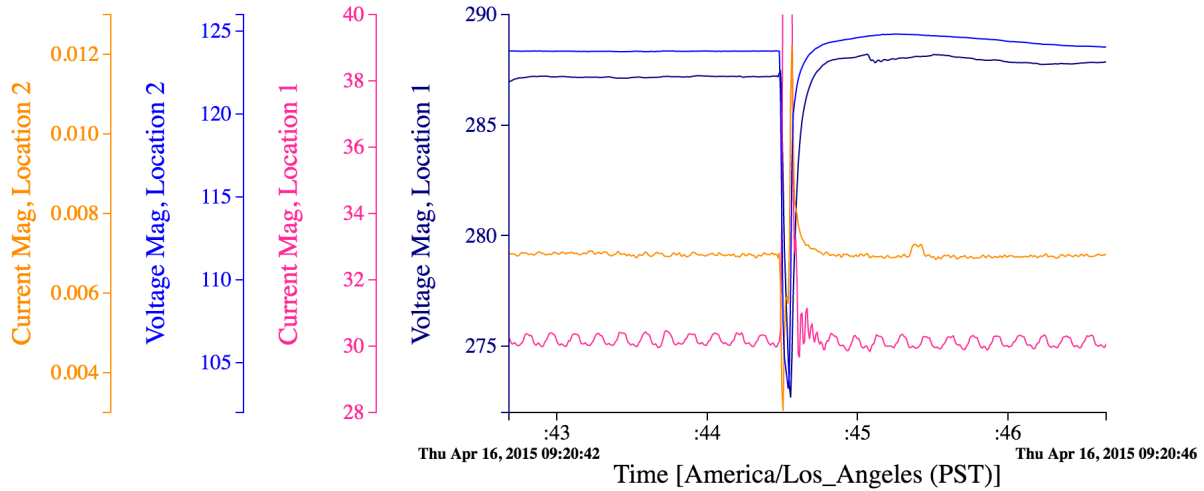

Figure 5. Sample event as seen on the PQube3 plotter (shown for one phase only). a voltage diturbance originating from the transmission system results in a current transient and the subsequent loss of some load. Location 1 is the PSL Inc in ALmedea, CA and Location 2 is the University of California campus in Berkeley. The data is available for public on [13].

## VI. APPLICATIONS FOR µPMU DATA

A broad spectrum of potential distribution system applications could hypothetically be supported by µPMU data (or, in some cases, by conventional PMU data), as has been noted in the literature [14, 15]. In the course of our research, we have explored a number of these applications, but there are many others that will require future work. The published report [16] is a comprehensive list of µPMU applications in distribution networks diagnostic applications.

It is also important to distinguish diagnostic from control applications. Diagnostic applications help operators better understand the present or past condition of the distribution system, while control applications inform specific actions to be taken, likely by automated systems, in more or less "real-time." Our research has only incorporated field data into diagnostic applications, but we are also exploring some control applications[17].

### A. Diagnostic Applications

Diagnostic applications that have been explored as part of this project include:

#### 1) Phase identification

Electric utilities often have limited or unreliable information about loads connected to three phase distribution lines. Moreover, phases can be switched during to restoration, reconfiguration and maintenance activities, which happen often in distribution networks. Such changes are not always tracked [18]. Correct phase labeling is crucial in order to avoid excessive losses or reduced life cycle of network components as a result of imbalanced loads or insufficient harmonic mitigation. Phase mislabeling is also a major source of error in diagnostic processes such as topology detection, state estimation, and fault location [19]. Phase identification is a nontrivial subset of topology detection that can be performed with substantially lower accuracy in phasor magnitude and angle. Matching phases between relatively nearby points is essentially a search for constant offsets of multiples of 30 degrees due to delta-wye transformers that may be between points. This can realistically be done with angular resolution as low as 1 degree. Based on available µPMU data, we have developed a novel phase identification method for distribution networks, where phases can be severely unbalanced and unlabeled. A key fact is that time-series voltage phasors taken from a distribution network show specific patterns regarding connected phases at measurement points. The algorithm is based on analyzing cross correlations over voltage magnitudes along with phase angle differences on two candidate phases to be matched. If two measurement points are on the same phase, large positive voltage magnitude correlations and small voltage angle differences should be observed [20].

#### 2) Topology status verification

Topology Detection makes explicit the open/closed status of switches at known locations. Knowledge of the network topology is essential to inform safe operations and accurate estimation of the system state. It might be assumed that switch statuses, and thus the network topology, are known prior to μPMU measurements, i.e. if switches are instrumented with the Supervisory Control And Data Acquisition (SCADA), but in practice this information is usually spotty and outdated. Although connectivity is not directly sensed by μPMUs, it can be inferred from the phase angle difference between points on opposite sides of a switch.

We have developed model-based and model-less approaches to detect the topology of distribution networks based on the analysis of time series μPMU measurements. For model-less approaches the key fact is that time-series data taken from a dynamic system show specific patterns regarding state transitions, such as opening or closing of switches, as a kind of signature from each topology change [21, 22]. The model-based approach is a voting-based algorithm that looks for the minimal difference between measured and calculated voltage angle or voltage magnitude to indicate the actual topology [23].

*3) State estimation*

State estimation combines knowledge of system topology and steady-state behavior, i.e. voltages and currents or real and reactive power flows. The objective of state estimation is to identify the steady-state voltage magnitudes and angles at each bus in a network, which completely characterizes the operating state of the system – meaning the real and reactive power flows on every link, as well as power injected into or withdrawn from each bus. State estimation is generally more difficult for distribution than for transmission systems, for several reasons: 1) Distribution systems are harder to model, owing to untransposed lines with phase imbalances, small X/R ratios, and large numbers of connecting load points; and 2) distribution systems present a high-dimensional mathematical problem with few physical measurements compared to the number of nodes, and less redundancy from Kirchhoff's laws. Data from μPMUs could ease these difficulties by directly feeding state variables (voltage angle and magnitude) into a Distributed State Estimator (DSE), which in turn may provide information to a Distribution Management System (DMS).

We have developed a Bayesian linear estimator based on a linear approximation of the power flow equations for distribution networks, which is computationally more efficient than standard nonlinear weighted least squares (WLS) estimators. We showed via numerical simulations that the developed linear state estimator performs similarly to the standard WLS estimator on a small distribution network [24].

*4) Model Validation and lessons learned from an actual installation*

Most utilities' distribution network models are based on geographical information system (GIS) data. GIS data is a series of interconnected features and layers documenting coordinates of key equipment and topology. The accuracy of the distribution model is inherently based upon the GIS systems and available information about network components and their parameters [12]. The ability to confirm or update utility models of distribution equipment has been one area of study for this project. μPMUs were installed and evaluated in several simple test configurations: on either side of a pair of single, three-phase lines; on either side of a transformer; and on either side of a three-phase underground cable. From the measurements returned by each set of μPMUs, our research group attempted to recreate the impedance values of the components connecting them.

In the first two cases, the overhead lines and the transformer, simple methods designed to check the overall feasibility of impedance estimation worked well. The process starts with the conventional matrix equations for voltage and current relations across a three-phase line and a transformer.

Voltage Drop across Two Three-Phase Lines:

$$[VLN_{1ABC}] - [VLN_{2ABC}] = [Z_{1ABC}][I_{1ABC}] - [Z_{2ABC}][I_{2ABC}] \qquad (2)$$

Where *VLN* and *I* variables are 3x1 vectors of voltage and current measurements at end nodes "1" and "2," and *Z* parameters are symmetric 3x3 impedance matrices.

Voltage Drop across Delta-Grounded Wye Transformer:

$$[A_t][VLN_{ABC}] - [VLG_{abc}] = [Z_{abc}][I_{abc}] \qquad (3)$$

Here, *Z* is a diagonal matrix, and *VLN* and *VLG* are line-to-neutral and line-to-ground voltages, respectively. Capital subscripts refer to the high-voltage side of the transformer while lowercase refer to the low.

$A_t$ in (3) is given by

$$[A_t] = \frac{1}{n_t} \begin{bmatrix} 1 & 0 & -1 \\ -1 & 1 & 0 \\ 0 & -1 & 1 \end{bmatrix} \quad (4)$$

$n_t$ being the ratio of the rated line-to-line voltage on the high side and the line-to-neutral voltage on the low side.

Taking the two equations (2) and (3), we then rearrange their terms and solve them for the elements of the Z matrices using ordinary least squares. The results compared reasonably well to the expected impedances from the utility models: our transformer impedance estimations were within 15% of the modeled values, as measured by Z matrix norms. Our line impedance estimations were within 13% by the same metric. This is an encouraging initial result, but we are in the process of developing and testing methods that we expect will significantly increase our accuracy by taking into account instrumentation transformer error.

There are points, though, beyond which that transformer error becomes significant enough that we have not been able to obtain even an approximate value of impedance. In the case of the underground cable, we have brought more sophisticated methods than ordinary least squares to bear, including constrained least squares and gradient and coordinate descent algorithms. However, every attempt to characterize the underground cable's impedance has failed, either returning estimations that are physically unreasonable or failing to converge.

These difficulties are likely due to the lack of cross-phase excitation in the measurement data for the underground cable. Because the cable itself is short and lightly loaded, the currents flowing through each of the three phases are not distinct enough from one another to be analyzed through general measurement noise. This means that there will be a lower limit on data "distinction" that must be achieved before our impedance estimation methods are able to function. Finding that threshold may be a subject of future research.

*5) Fault location*

The goal is to infer the actual geographical location of a fault on a distribution feeder to within a small circuit section (compared to the distance between protective devices) by using recorded measurements of voltage angle before and during the fault, and interpreting these in the context of a circuit model. It is critical for a resilient grid operation with fast service restoration after an outage. Algorithms exist for locating faults through proper analysis of monitored data, but the quality of available measurements on distribution circuits is often insufficient to support them. The addition of μPMU data to conventional substation current measurement devices improves upon traditional impedance methods by measuring the behavior of the system on both sides of the fault, making the fault location estimate more robust to the distance of the fault from the substation, variations in system conditions, and uncertainties in system models.

Using μPMU data, we have a phasor-based fault location algorithm for distribution networks [25]. Rather than requiring many specialized line sensors to enable fault location, the proposed approach leverages the μPMU data. The accuracy of existing fault location methods are dependent either on dense deployments of line sensors or unrealistically accurate models of distribution networks. The proposed algorithm uses pre- and post-fault voltage phasor values at the substation and remote μPMUs, as well as current measurements at the substation, in order to pinpoint a fault in short time.

*6) Event detection*

Events of interest in power distribution networks are sinusoidal or non-sinusoidal transients in voltage and current waveforms that may be caused by faults, topology changes, load behavior and source dynamics. These events include, but are not limited to, voltage sags, voltage swells, fault currents, voltage oscillations, and frequency oscillations. For the sake of power systems reliability and stability, it is crucial to monitor the operating states in real time and detect anomalies quickly as to avert disturbances and disruptions.

In literature, the model-based event detection has been very popular and successful in many applications. However, it is prone to overwhelming system randomness and dynamics in our context owing to the high time resolution and dimensionality of our measurements. In particular, model-based approaches rely heavily on correctness of the dynamical model of the system, as well as system analytic tools such as real time state estimators, parameter estimation, parity equations etc. Their limitations are obvious: (1) the dynamics of a system may be hard to specify in many cases and (2) they have nonlinear structures. Therefore, the focus of the present work is to detect abnormal events based on uPMU measurements by considering only the empirical data. The data-driven approach uses methods of machine learning/pattern recognition to conduct statistical inference or decision making on available system measurements, and is receiving increasing attention in both application and research domains. We developed a machine learning-based discriminative method for anomaly detection in uPMU measurements. It is a kernel Principle Component Analysis (kPCA) that builds statistical models for nominal states and detects possible anomalies in order to build tight boundaries describing the support of the normal distribution [26]. (see figure 6, the data available for public on [13])

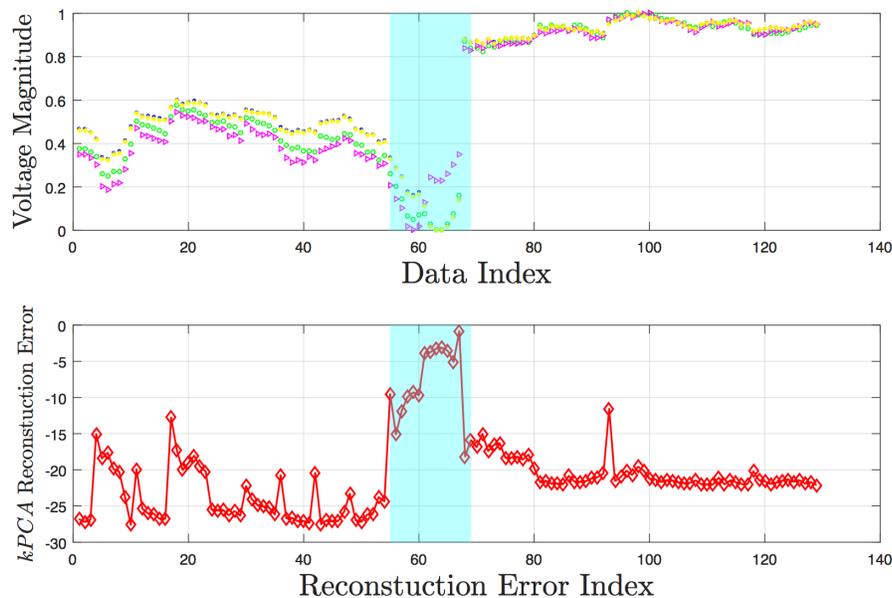

Figure 6. Event Detection based on kPCA, above figure is voltage sag event observed by a uPMU. The bottom figure is the kPCA reconstruction errors for the testing sequence.

7) *Oscillation detection*

Subsynchronous oscillations are known to exist on transmission systems, and higher-frequency oscillations could conceivably occur on distribution systems, unobserved by conventional instrumentation. These could be the result of power exchange between distributed energy resources or resonance phenomena on the circuit. Low-frequency modes of oscillation, though normally well damped, constrain a.c. transmission paths and can grow destructive if underdamped. It took synchrophasors to recognize their existence, and effective control methods are still in development. Observation of oscillation modes on the island of Maui, measured at transmission voltage but across a small geographic scale (tens of miles), suggests that future distribution systems with high penetrations of solar and wind generation could also experience oscillation issues [15].

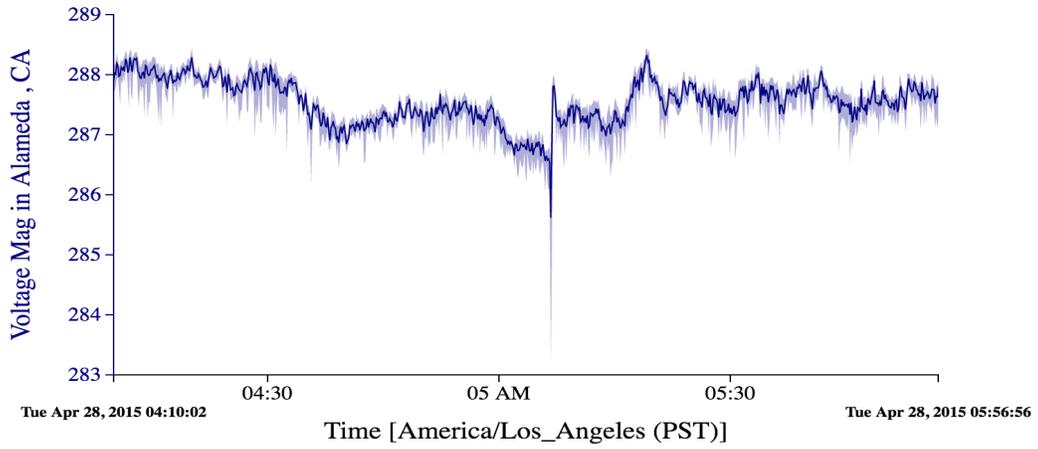

(a)

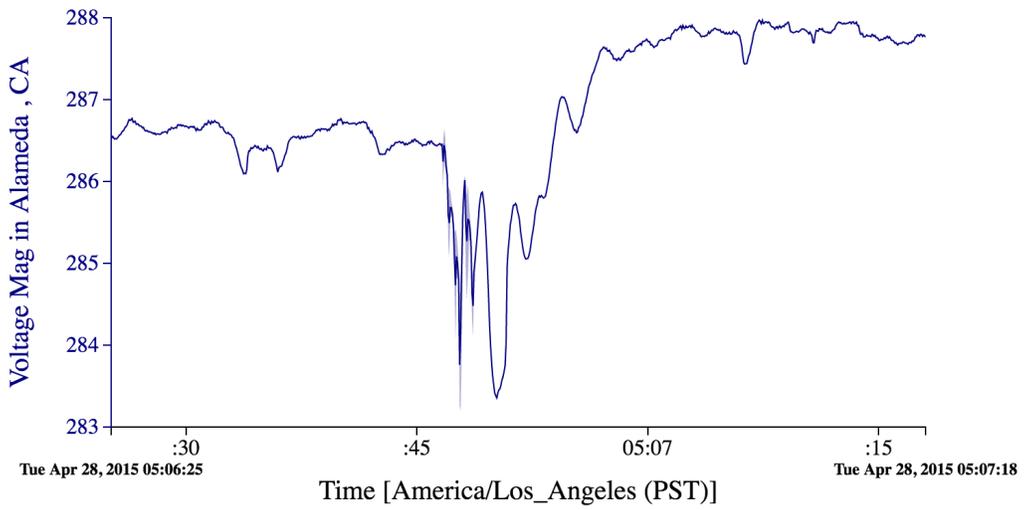

(b)

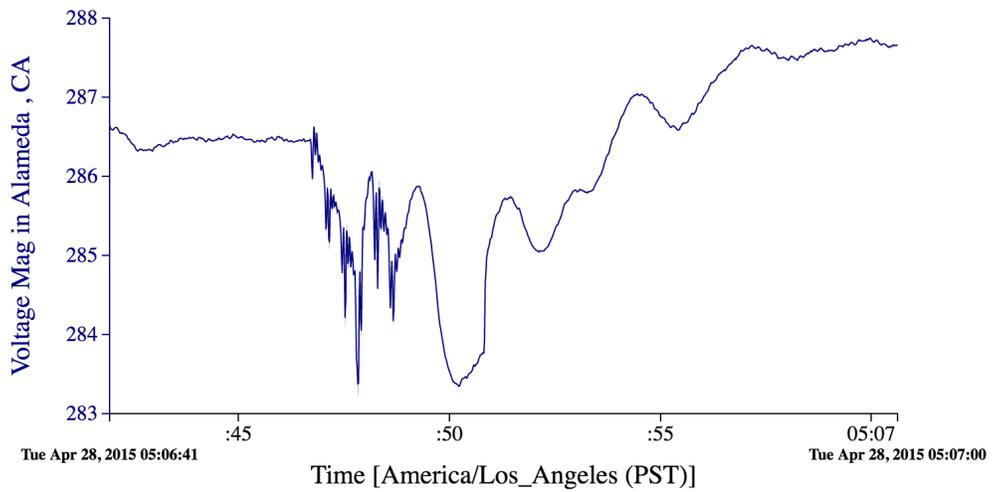

(c)

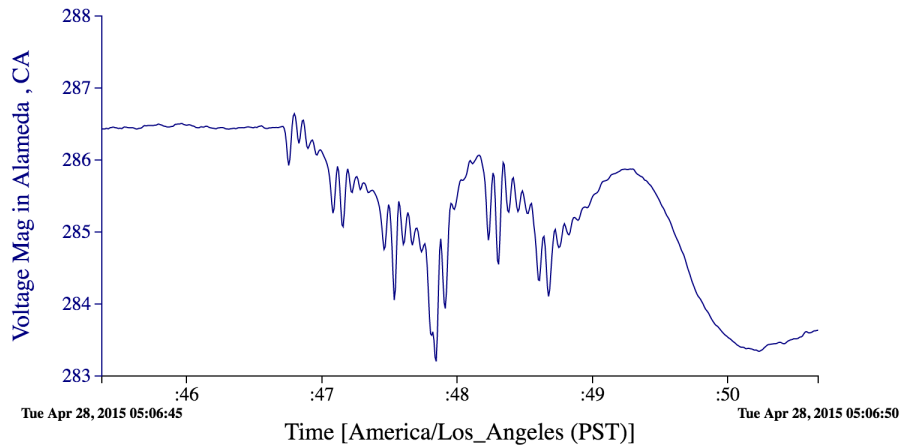

**(d)**

Figure 7. Voltage Magnitude during a protection trip on the Pacific Direct Current Intertie (PDCI) on April 28 2015, 05:06:46 am PST recorded by a PQube3 installed at the PSL Inc at Alameda, CA. PQube3 data for some selected devices are available for public on [13] and [7].
  a. Snapshot of the voltage magnitude with an hour time-scale.
  b. Snapshot of the same event with the 5 minutes time-scale
  c. Snapshot of the same event with the 15 seconds time-scale
  d. Snapshot of the same event with one second time-scale

Transmission system models did not predict oscillations, nor do distribution system models; the only way to find out if any oscillations exist – and if so, to characterize them – is to look. From the PSL Inc PQube3 installation, we have observed a fault on the Pacific Direct Current Intertie (PDCI) in the Western Electricity Coordinating Council (WECC). The data for some selected Pqube3 devices are available for public on [13] and [7]. See Figure 7-a, a frequency drop happened at 5:06' that was caused by a generation loss and it lasted for more almost 25 minutes. Figure 7-b shows the frequency pick up after the fault. Figure 7-c depicts the detailed oscillation in the few seconds after the fault. Finally, figure 7-d shows oscillations preceding the loss of generation that might be caused by interactions inside the distribution network following the transmission level event. The data from multiple μPMUs can help us to observe, analyze and understand events such as these in distribution networks. If the data is of high enough quality, the operator can pinpoint the source of events. This figure also illustrates the capability of user interface for μPMU data visualization.

Other diagnostic applications are listed as follows (see table 1):

*8) High-impedance fault detection*

The objective is to recognize the dangerous condition where an object such as a downed power line makes an unintentional connection with the ground, but does not draw sufficient current to trip a protective device (since it mimics a legitimate load).

*9) Unintentional island detection*

The objective is to quickly and reliably recognize a potentially unsafe situation where a set of distributed generators (DG) and loads have separated from the grid but continue to energize their local portion of the network. Today's inverters have very reliable anti-islanding protection. However, with greater penetration of diverse distributed resources and more complex dynamics on distribution circuits, it may become increasingly difficult to distinguish fault events from other abnormal conditions where it is desirable to keep DG online (for example, low-voltage ride-through).

*10) Reverse power flow detection*

The goal is to identify, or rather anticipate, when power flows in reverse direction on a radial distribution feeder. While reverse flow may be unproblematic in some situations, its significance depends on the type of protection system design used, and whether the coordination of protective devices could be compromised under reverse flow conditions. Voltage regulation may also be impacted by reverse flow. Moreover, the resultant harmonics from DG sources with reverse power flow can cause operational challenges. We have observed reverse power flow on distribution circuits with our μPMUs installed, but have not yet begun work on modeling or anticipating the circumstances under which it arises.

*11) Characterization of distributed generation*

The goal is to qualify and quantify the behavior of inverters in relation to stabilizing system a.c. frequency and damping disturbances in power angle or frequency. Specifically, this means observation of inverter real and reactive power output at very small time scales relative to line voltage, frequency and angle, with particular emphasis on the response to abnormal and transient conditions.

*12) FIDVR identification and risk detection*

Fault-induced delayed voltage recovery (FIDVR) is an unstable operating condition that results from the interaction of stalled air conditioners with capacitor bank controls [27]. Anticipating FIDVR before it occurs would hinge on identifying in near real-time the varying contribution to total customer load from devices such as single-phase induction motors in residential and small commercial air conditioners that pose an increased risk.

*13) Unmasking loads from net metered DG*

The goal is to infer the amount of load being offset by distributed generators (DG) behind a net meter through measurements and correlated data obtained outside the customer's premises. Estimating the real-time levels of renewable generation versus loads would allow for better anticipation of changes in the net load, by separately forecasting the load and generation, and for assessing the system's risk exposure to sudden generation loss. At the aggregate level, this information is of interest to system operators for evaluating stability margins and damping levels in the system.

Table 1. Summary of Attributes of μPMU Applications

| Diagnostic Application | Competing conventional strategies | Likely advantage of voltage angle | Likely technical challenges |
|---|---|---|---|
| **Unintentional island detection** | Automatic Transfer Switch | faster, greater sensitivity and selectivity | communication latency |
| **Oscillation detection** | none | Unique and crucial | Potential Transformer and Current Trasnformer errors |
| **Reverse power flow detection** | Line sensors, directional relays | may extrapolate to locations not directly monitored | μPMU placement |
| **FIDVR detection** | detected with V mag | more accurate, with phase angle mesaurement | easy |
| **Fault location** | various | better accuracy using voltage phasor values | communication latency |
| **High-impedance fault detection** | various, difficult | better sensitivity and selectivity using voltage phasor values | communication latency |
| **Topology detection** | direct SCADA on switches | fewer measurement points, higher accuracy using timeseries phasor data | μPMU placement |
| **State estimation** | computation based on V mag measurements | linera state estimatiin, higher accuracy | μPMU placement |
| **Dynamic circuit monitoring** | high-resolution Power Quality (PQ) instruments, none for δ | uniquely capture oscillations, damping | data mining for relevant phenomena |
| **Load and DG characterization** | limited observation with PQ instruments | uniquely capture dynamic behaviors | data mining, proximity to subject |
| **Unmasking load/DG** | none | unique | data mining, proximity to subject |

### B. Control Applications

Beyond enhanced diagnostic capabilities, synchrophasor data may enable more refined management and active control of distribution systems. Our group studied phasor-based controls by deriving linear relationships between the power injected, voltage magnitude, and voltage phase angle on a three-phase, unbalanced feeder [28]. These relationships could be leveraged in a number of possible control strategies:

#### 1) Volt-VAR Optimization

Present utility efforts aimed at Volt-Var Optimization (VVO) consider time scales on the order of tens of seconds or minutes. Legacy voltage regulation commonly operates with even less time resolution; hourly adjustments are common for capacitor banks. Since load tap changer (LTC) and voltage regulators have mechanical parts subject to wear and premature failure, it is considered undesirable to operate them more frequently. Although more sophisticated hardware can perform VVO-related operations on timescales of seconds, it is not obvious that significant additional benefits will result from higher temporally granular levels of control. Consequently, the relevant time step for VVO would be much greater than one second, likely closer to ten seconds or even a minute. A realistic performance target for steady-state VVO is <5 seconds from measurement to initiating the control action. As such, fast sampling rate, communications latency and control decision speed should be readily within the capabilities supported by the μPMU.

As decentralized approaches are typically sub-optimal in nature, it is not expected that standard deviations on the order of 0.01 degrees for voltage angle measurements , or $10^{-4}$ p.u. for voltage magnitude measurements, would impair control actions for VVO. A realistic performance target for voltage control would be on the order of $10^{-3}$ p.u., or 0.1 V on a 120 V base[28].

#### 2) Distrbuted Eneregy Resources (DER) Coordination

The process of dispatching controllable DERs can be conducted at many different time scales, depending on the objective. As mentioned earlier, academic research regarding coordination of DERs in distribution systems oftentimes involves casting the problem as a model-based centralized optimization program (similar to VVO). As such, these strategies depend on accurate models and implicit communications requirements. In addition, many of the state of the art approaches for asset coordination in distribution systems are based on semi-definite convex relaxations of the optimal power flow problem. Although these problems are convex, tightness of the relaxed problem (to our knowledge) has yet to be proven, and solving large semi-definite programs may be time consuming. For particular situations, the computation time required to determine control strategies for all controllable DER assets based on semi-definite programming may be on the order of several seconds. Therefore, it is desirable to explore approaches for DER coordination that are decentralized. It is envisioned that such strategies will place increased emphasis on the utilization of time synchronized phasor measurement data for decision- making.

Similar to VVO, the decentralized approach to DER coordination is expected to be sub-optimal in the mathematical sense: the emphasis is not on obtaining ideal results, which would require a level of model detail that we deem unrealistic, but to maintain quantities within acceptable bounds. Therefore, it is not expected that standard deviations in voltage angle measurements on the order of 0.01 degrees would adversely impact control system performance for DER coordination.

#### 3) Microgrid Control

Generation and load within a power island can be balanced through conventional frequency regulation techniques, but explicit phase angle measurement may prove to be a more versatile indicator. In particular, angle data may provide for more robust and flexible islanding and re-synchronization of microgrids. A convenient property of PMU data for matching frequency and phase angle is that the measurements on either side need not be at the identical location as the physical switch between the island and the grid. A self-synchronizing island that matches its voltage phase angle to the core grid could be arbitrarily disconnected or paralleled, without even momentary interruption of load. Initial tests of such a strategy with angle-based control of a single generator were found to enable smooth transitions under continuous load with minimal discernible transient effects [29].

Comparison of angle difference between a microgrid or local resource cluster and a suitably chosen point on the core grid could enable the cluster to provide ancillary services as needed, and as determined by direct, physical measurement of system stress rather than a price signal – for example, by adjusting power imports or exports to keep the phase angle difference within a predetermined limit. A variation of this approach, known as angle-constrained active management (ACAM), has been demonstrated in a limited setting with two wind generators on a radial distribution circuit[30]. In combination, these capabilities imply the possibility of distributed resources able to smoothly transition between connected and islanded states, and capable of providing either or both local power quality & reliability services, and support services to the core grid, as desired at any given time.

The control speed requirement is based on the amount of time for which the steady-state angle and power estimate across the Point of Common Coupling (PCC) can be assumed to remain valid. Note that there is a substantial difference in the speed demands of this application as compared to those that optimize voltages and power flows: here, an excessive delay could result in a physical mismatch that causes actual damage, rather than a minor deviation from a desired operating point. If either the micro- or macro-grid is undergoing significant dynamic changes, this could be less than a cycle. However, it seems reasonable to presume that the PCC would only be closed after verifying that both sides are in a stable, steady state. It then becomes fair to assume that the phase angle difference and power transfer measurement remains valid for several cycles. Taking six cycles as a reference point, this would require latency on the order of 0.1 seconds for the microgrid resynchronization application. Within that time frame, it will be necessary to compare system states on either side of the PCC to determine if reconnecting is feasible. Therefore, the requirements for µPMU measurement resolution would be comparable to those in the DER Coordination application. Table 2 presents the data requirements for control applications.

Table 2. Data Requirements for Control application requirements [28].

| Application | Sampling Rate (per second) | Latency | Speed of Control | Spatial Coordination | Measurement Resolution |
|---|---|---|---|---|---|
| VVO | < 120 | > 200 ms | Slow | Low | Commensurate |
| DER Coordination | < 120 | > 200 ms | Slow | Low | Commensurate |
| Microgrid Control | < 120 | > 200 ms | Slow | Medium-High | Commensurate |

## VII. µPMU DATA SPECIFICATIONS

The lack of high-quality measurement data in distribution networks is a compelling motivation for using advanced measurement data from accurate, high resolution devices such as µPMUs. Applications will differ in the quantity, quality, and timeliness of measurement data they require. From a practical point of view, the voltage and current phasor measurement are subject to error introduced by potential and current transformers, or any loaded transformers between the µPMU and the voltage quantity of interest. This section distinguishes key dimensions of data requirements and discusses the ability of the µPMU device itself to meet relevant criteria for each application. An extensive discussion on µPMU data requirements and measurement accuracy is available in our publications [16, 31].

Table 3. Expected Data requirements for different classes of µPMU applications

| | Sampling rate (per cycle) | Angle Resolution (milli-deg) | Spatial Resolution (placement) | Data volume (Bandwidth) | Commun Speed |
|---|---|---|---|---|---|
| **Steady-state circuit behavior** | 1 | 10-100 | Sparse | Medium but continuous | usually low |
| **Dynamic circuit behavior** | 2-512 | 10-100 | Dense | High but could be intermittent | usually high |

*A. Sampling Rate*

While the µPMU is capable of 512 samples per cycle (approx. 30 kHz) for power quality measurements (e.g. harmonic content), the focus is on time-synchronized measurement of the fundamental voltage and current phasors. It is only meaningful over time periods of at least one cycle. In synchrophasor mode, the voltage and current magnitude and phase angle data are communicated by the µPMU twice per cycle, or 120 times per second. The 120 samples per second is more than enough for steady state applications such state estimation and topology detection. For dynamic and transient applications, such as fault or islanding detection and voltage volatility measurement of photovoltaic sources, the μPMU can be run on power quality mode. The distinct requirements of data in terms of both data reporting and device placement is shown in Table 3.

*B. Resolution*

Angular resolution refers to the minimum difference between reference points (e.g. zero crossing or peak) on the voltage waveform at different measurement locations, translated into units of angle, that can be discriminated. This hinges on the precision and accuracy of the time stamp assigned to the voltage measurement at each μPMU. It is worth noting that high sampling rates are not important for obtaining angular resolution, which depends rather on the accuracy of the GPS time stamp. The μPMU is capable of measuring phase angle with a resolution of 0.01 degrees or $10^{-4}$ rad [4]. The experimental tests of voltage magnitude measurements showed that the μPMU is capable of measuring rms voltage with a standard deviation on the order of 0.02 Vrms (or $1.7 \times 10^{-4}$ p.u. on a 120-V base)[4]. This level of resolution can observe the explicit voltage variations that are of practical interest on many application in distribution network (on the order of 0.01 p.u.) [16].

*C. Spatial Data Characteristics*

Spatial coordination refers to the degree to which μPMUs must exchange information with one another or a data concentrator unit. In order to utilize voltage phase angle information, a comparison must be made between a reference μPMU and at least the second device in a different location. In the context of diagnostic applications, we consider instrumenting distribution circuits with multiple μPMUs. For control applications, even the complexity of collecting and interpreting data from represents a disadvantage in terms of latency, processing time, and additional information required (such as network models) [28].

*D. Data transfer rate*

While all measurements are stored in a circular buffer on the μPMU device, options for exporting data in real-time via ethernet or wireless communications range from firehose mode to arbitrarily sparse data selections at long intervals, and reports triggered by exceptional measurements. In the developed DISTIL tool for μPMU data mining and analysis, raw data is streamed directly into the database via the chunk loader based on the available bandwidth for data transmission network [10]. Usually, the most data-hungry applications will determine the settings for a given μPMU.

*E. Application specific data requirment, State Estimatuion*

The performance of a state estimation algorithm in practice will depend on the quality of the input data, the network model, and the number of μPMUs. Naturally, the fewer nodes that are instrumented with μPMUs, the better the measurement needs to be in order to yield a good state estimate. As it is mentioned earlier, the total vector error in μPMU data will be dominated by errors introduced by potential and current transformers. Table 4 shows the data requirement for state estimation application. The accuracy is measured by the total vector error (TVE).

Table 4. Expected Data requirements for State Estimation Application

| State Estimation Use Case | Required Accuracy (TVE) | Acceptance Latency |
|---|---|---|
| Emergency Alarms | ~ 1-5% | 5-15 min |
| Avoid Constraints Violations | ~ 0.5% | 5 min |
| Improve System Efficiency | ~ 0.5% | 30 sec |

*F. Application specific data requirment, Topolgy Detection*

For topology detection, the essential performance criterion is the confidence level at which a given device (say, circuit breaker or sectionalizing switch) can be identified as open or closed, or conversely, the error rate of the topology detection algorithm in reporting the status of an individual device. For topology detection data requirements, three types of use cases are considered that call for different degrees of confidence: 1. providing positive identification of switch status as the sole source for safety-critical decisions, 2. corroborating independent switch status information (from SCADA or field crew records) to increase the confidence level in safety-critical operational decisions, and 3. support state estimation (e.g. flag topology errors). The ideal topology detection tool aims to detect switch status with high confidence in less than 15 minutes, so as to be able to match topology with load curtailment as it shows in the smart meter data. Table 5 depicts µPMU data requirements for topology detection in the three use cases.

Table 5. Expected Data requirements for State Estimation Application

| State Estimation Use Case | Required Accuracy (TVE) | Acceptance Latency |
|---|---|---|
| Switch Status Identification | ~ 0.000 % | 5-15 min |
| Corroborate field crew or SACDA information | ~ 1% | 5 min |
| Support State Estimation | ~ 5% | 1 min |

## VIII. CONCLUSION

Affordable, high-resolution measurement of voltage phase angle may offer significant new options for actively managing distribution systems with diverse resources and growing complexity. The initial groundwork that has been laid by our efforts in model validation and development of mathematical frameworks for control is an important step in exploring those options, but there are a great number of phasor-based analysis and control applications still to be explored.

## IX. ACKNOWLEDGMENTS

I thank our colleagues on the project team especially Prof. Alexandra von Meier and Mr. Kyle Bradly.## X. REFERENCES

[1] A. von Meier and R. Arghandeh, "Chapter 34 - Every Moment Counts: Synchrophasors for Distribution Networks with Variable Resources," in *Renewable Energy Integration*, L. E. Jones, Ed., ed Boston: Academic Press, 2014, pp. 429-438.
[2] R. Arghandeh, A. Onen, J. Jung, D. Cheng, R. Broadwater, and V. Centeno, "Harmonic Impact Study For Distributed Energy Resources Integrated into Power Distribution Networks," in *American Society of Mechanical Engineers Power Conference 2013, (ASME Power2013), Boston, MA, USA*, 2013, p. 8.
[3] A. von Meier, D. Culler, A. McEachen, and R. Arghandeh, "Micro-synchrophasors for distribution systems," in *IEEE PES Innovative Smart Grid Technologies Conference (ISGT), 2014*, 2014, pp. 1-5.
[4] "PQube Specifications," PSL, Ed., ed. USA: PSL, 2015.
[5] A. G. Phadke and J. S. Thorp, *Synchronized phasor measurements and their applications*: Springer Science & Business Media, 2008.
[6] J. De La Ree, V. Centeno, J. S. Thorp, and A. G. Phadke, "Synchronized Phasor Measurement Applications in Power Systems," *Smart Grid, IEEE Transactions on,* vol. 1, pp. 20-27, 2010.
[7] PSL. (2015). *Power Sensors Limited, PQube3* Available: http://www.powersensorsltd.com/PQube
[8] H. Gharavi and H. Bin, "Scalable Synchrophasors Communication Network Design and Implementation for Real-Time Distributed Generation Grid," *Smart Grid, IEEE Transactions on,* vol. 6, pp. 2539-2550, 2015.
[9] M. Patel, S. Aivaliotis, and E. Ellen, "Real-time application of synchrophasors for improving reliability," *NERC Report, Oct,* 2010.
[10] S. K. Michael P Andersen, Connor Brooks, Alexandra von Meier and David E. Culler, "DISTIL Design and Implemention of a Scalable Synchrophasor Data Processing System," presented at the ACM sigcomm 2015, 2015.
[11] T. J. Overbye and J. D. Weber, "Visualization of power system data," in *System Sciences, 2000. Proceedings of the 33rd Annual Hawaii International Conference on*, 2000, p. 7 pp.
[12] E. M. Stewart, S. Kiliccote, C. M. Shand, A. W. McMorran, R. Arghandeh, and A. von Meier, "Addressing the Challenges for Integrating Micro-Synchrophasor Data with Operational System Applications," in *IEEE PES General Meeting 2014*, 2014, pp. 1-5.
[13] "Mr. Plotter, https://plot.upmu.org/," in *University of California Berkeley and PSL Inc*, ed. https://plot.upmu.org/, 2015.17